\documentclass[letterpaper,compsoc,twoside]{IEEEtran}
\usepackage{fixltx2e} 
\usepackage{cmap} 
\usepackage{ifthen}
\usepackage[T1]{fontenc}
\usepackage[utf8]{inputenc}
\usepackage{amsmath}

\usepackage[font={small,it},labelfont=bf]{caption}
\usepackage{float}

\usepackage{scipy}
\makeatletter
\def\PY@reset{\let\PY@it=\relax \let\PY@bf=\relax%
    \let\PY@ul=\relax \let\PY@tc=\relax%
    \let\PY@bc=\relax \let\PY@ff=\relax}
\def\PY@tok#1{\csname PY@tok@#1\endcsname}
\def\PY@toks#1+{\ifx\relax#1\empty\else%
    \PY@tok{#1}\expandafter\PY@toks\fi}
\def\PY@do#1{\PY@bc{\PY@tc{\PY@ul{%
    \PY@it{\PY@bf{\PY@ff{#1}}}}}}}
\def\PY#1#2{\PY@reset\PY@toks#1+\relax+\PY@do{#2}}

\expandafter\def\csname PY@tok@gd\endcsname{\def\PY@tc##1{\textcolor[rgb]{0.63,0.00,0.00}{##1}}}
\expandafter\def\csname PY@tok@gu\endcsname{\let\PY@bf=\textbf\def\PY@tc##1{\textcolor[rgb]{0.50,0.00,0.50}{##1}}}
\expandafter\def\csname PY@tok@gt\endcsname{\def\PY@tc##1{\textcolor[rgb]{0.00,0.27,0.87}{##1}}}
\expandafter\def\csname PY@tok@gs\endcsname{\let\PY@bf=\textbf}
\expandafter\def\csname PY@tok@gr\endcsname{\def\PY@tc##1{\textcolor[rgb]{1.00,0.00,0.00}{##1}}}
\expandafter\def\csname PY@tok@cm\endcsname{\let\PY@it=\textit\def\PY@tc##1{\textcolor[rgb]{0.25,0.50,0.56}{##1}}}
\expandafter\def\csname PY@tok@vg\endcsname{\def\PY@tc##1{\textcolor[rgb]{0.73,0.38,0.84}{##1}}}
\expandafter\def\csname PY@tok@m\endcsname{\def\PY@tc##1{\textcolor[rgb]{0.13,0.50,0.31}{##1}}}
\expandafter\def\csname PY@tok@mh\endcsname{\def\PY@tc##1{\textcolor[rgb]{0.13,0.50,0.31}{##1}}}
\expandafter\def\csname PY@tok@cs\endcsname{\def\PY@tc##1{\textcolor[rgb]{0.25,0.50,0.56}{##1}}\def\PY@bc##1{\setlength{\fboxsep}{0pt}\colorbox[rgb]{1.00,0.94,0.94}{\strut ##1}}}
\expandafter\def\csname PY@tok@ge\endcsname{\let\PY@it=\textit}
\expandafter\def\csname PY@tok@vc\endcsname{\def\PY@tc##1{\textcolor[rgb]{0.73,0.38,0.84}{##1}}}
\expandafter\def\csname PY@tok@il\endcsname{\def\PY@tc##1{\textcolor[rgb]{0.13,0.50,0.31}{##1}}}
\expandafter\def\csname PY@tok@go\endcsname{\def\PY@tc##1{\textcolor[rgb]{0.20,0.20,0.20}{##1}}}
\expandafter\def\csname PY@tok@cp\endcsname{\def\PY@tc##1{\textcolor[rgb]{0.00,0.44,0.13}{##1}}}
\expandafter\def\csname PY@tok@gi\endcsname{\def\PY@tc##1{\textcolor[rgb]{0.00,0.63,0.00}{##1}}}
\expandafter\def\csname PY@tok@gh\endcsname{\let\PY@bf=\textbf\def\PY@tc##1{\textcolor[rgb]{0.00,0.00,0.50}{##1}}}
\expandafter\def\csname PY@tok@ni\endcsname{\let\PY@bf=\textbf\def\PY@tc##1{\textcolor[rgb]{0.84,0.33,0.22}{##1}}}
\expandafter\def\csname PY@tok@nl\endcsname{\let\PY@bf=\textbf\def\PY@tc##1{\textcolor[rgb]{0.00,0.13,0.44}{##1}}}
\expandafter\def\csname PY@tok@nn\endcsname{\let\PY@bf=\textbf\def\PY@tc##1{\textcolor[rgb]{0.05,0.52,0.71}{##1}}}
\expandafter\def\csname PY@tok@no\endcsname{\def\PY@tc##1{\textcolor[rgb]{0.38,0.68,0.84}{##1}}}
\expandafter\def\csname PY@tok@na\endcsname{\def\PY@tc##1{\textcolor[rgb]{0.25,0.44,0.63}{##1}}}
\expandafter\def\csname PY@tok@nb\endcsname{\def\PY@tc##1{\textcolor[rgb]{0.00,0.44,0.13}{##1}}}
\expandafter\def\csname PY@tok@nc\endcsname{\let\PY@bf=\textbf\def\PY@tc##1{\textcolor[rgb]{0.05,0.52,0.71}{##1}}}
\expandafter\def\csname PY@tok@nd\endcsname{\let\PY@bf=\textbf\def\PY@tc##1{\textcolor[rgb]{0.33,0.33,0.33}{##1}}}
\expandafter\def\csname PY@tok@ne\endcsname{\def\PY@tc##1{\textcolor[rgb]{0.00,0.44,0.13}{##1}}}
\expandafter\def\csname PY@tok@nf\endcsname{\def\PY@tc##1{\textcolor[rgb]{0.02,0.16,0.49}{##1}}}
\expandafter\def\csname PY@tok@si\endcsname{\let\PY@it=\textit\def\PY@tc##1{\textcolor[rgb]{0.44,0.63,0.82}{##1}}}
\expandafter\def\csname PY@tok@s2\endcsname{\def\PY@tc##1{\textcolor[rgb]{0.25,0.44,0.63}{##1}}}
\expandafter\def\csname PY@tok@vi\endcsname{\def\PY@tc##1{\textcolor[rgb]{0.73,0.38,0.84}{##1}}}
\expandafter\def\csname PY@tok@nt\endcsname{\let\PY@bf=\textbf\def\PY@tc##1{\textcolor[rgb]{0.02,0.16,0.45}{##1}}}
\expandafter\def\csname PY@tok@nv\endcsname{\def\PY@tc##1{\textcolor[rgb]{0.73,0.38,0.84}{##1}}}
\expandafter\def\csname PY@tok@s1\endcsname{\def\PY@tc##1{\textcolor[rgb]{0.25,0.44,0.63}{##1}}}
\expandafter\def\csname PY@tok@gp\endcsname{\let\PY@bf=\textbf\def\PY@tc##1{\textcolor[rgb]{0.78,0.36,0.04}{##1}}}
\expandafter\def\csname PY@tok@sh\endcsname{\def\PY@tc##1{\textcolor[rgb]{0.25,0.44,0.63}{##1}}}
\expandafter\def\csname PY@tok@ow\endcsname{\let\PY@bf=\textbf\def\PY@tc##1{\textcolor[rgb]{0.00,0.44,0.13}{##1}}}
\expandafter\def\csname PY@tok@sx\endcsname{\def\PY@tc##1{\textcolor[rgb]{0.78,0.36,0.04}{##1}}}
\expandafter\def\csname PY@tok@bp\endcsname{\def\PY@tc##1{\textcolor[rgb]{0.00,0.44,0.13}{##1}}}
\expandafter\def\csname PY@tok@c1\endcsname{\let\PY@it=\textit\def\PY@tc##1{\textcolor[rgb]{0.25,0.50,0.56}{##1}}}
\expandafter\def\csname PY@tok@kc\endcsname{\let\PY@bf=\textbf\def\PY@tc##1{\textcolor[rgb]{0.00,0.44,0.13}{##1}}}
\expandafter\def\csname PY@tok@c\endcsname{\let\PY@it=\textit\def\PY@tc##1{\textcolor[rgb]{0.25,0.50,0.56}{##1}}}
\expandafter\def\csname PY@tok@mf\endcsname{\def\PY@tc##1{\textcolor[rgb]{0.13,0.50,0.31}{##1}}}
\expandafter\def\csname PY@tok@err\endcsname{\def\PY@bc##1{\setlength{\fboxsep}{0pt}\fcolorbox[rgb]{1.00,0.00,0.00}{1,1,1}{\strut ##1}}}
\expandafter\def\csname PY@tok@kd\endcsname{\let\PY@bf=\textbf\def\PY@tc##1{\textcolor[rgb]{0.00,0.44,0.13}{##1}}}
\expandafter\def\csname PY@tok@ss\endcsname{\def\PY@tc##1{\textcolor[rgb]{0.32,0.47,0.09}{##1}}}
\expandafter\def\csname PY@tok@sr\endcsname{\def\PY@tc##1{\textcolor[rgb]{0.14,0.33,0.53}{##1}}}
\expandafter\def\csname PY@tok@mo\endcsname{\def\PY@tc##1{\textcolor[rgb]{0.13,0.50,0.31}{##1}}}
\expandafter\def\csname PY@tok@mi\endcsname{\def\PY@tc##1{\textcolor[rgb]{0.13,0.50,0.31}{##1}}}
\expandafter\def\csname PY@tok@kn\endcsname{\let\PY@bf=\textbf\def\PY@tc##1{\textcolor[rgb]{0.00,0.44,0.13}{##1}}}
\expandafter\def\csname PY@tok@o\endcsname{\def\PY@tc##1{\textcolor[rgb]{0.40,0.40,0.40}{##1}}}
\expandafter\def\csname PY@tok@kr\endcsname{\let\PY@bf=\textbf\def\PY@tc##1{\textcolor[rgb]{0.00,0.44,0.13}{##1}}}
\expandafter\def\csname PY@tok@s\endcsname{\def\PY@tc##1{\textcolor[rgb]{0.25,0.44,0.63}{##1}}}
\expandafter\def\csname PY@tok@kp\endcsname{\def\PY@tc##1{\textcolor[rgb]{0.00,0.44,0.13}{##1}}}
\expandafter\def\csname PY@tok@w\endcsname{\def\PY@tc##1{\textcolor[rgb]{0.73,0.73,0.73}{##1}}}
\expandafter\def\csname PY@tok@kt\endcsname{\def\PY@tc##1{\textcolor[rgb]{0.56,0.13,0.00}{##1}}}
\expandafter\def\csname PY@tok@sc\endcsname{\def\PY@tc##1{\textcolor[rgb]{0.25,0.44,0.63}{##1}}}
\expandafter\def\csname PY@tok@sb\endcsname{\def\PY@tc##1{\textcolor[rgb]{0.25,0.44,0.63}{##1}}}
\expandafter\def\csname PY@tok@k\endcsname{\let\PY@bf=\textbf\def\PY@tc##1{\textcolor[rgb]{0.00,0.44,0.13}{##1}}}
\expandafter\def\csname PY@tok@se\endcsname{\let\PY@bf=\textbf\def\PY@tc##1{\textcolor[rgb]{0.25,0.44,0.63}{##1}}}
\expandafter\def\csname PY@tok@sd\endcsname{\let\PY@it=\textit\def\PY@tc##1{\textcolor[rgb]{0.25,0.44,0.63}{##1}}}

\def\PYZbs{\char`\\}
\def\PYZus{\char`\_}
\def\PYZob{\char`\{}
\def\PYZcb{\char`\}}

\def\PYZsh{\char`\#}

\def\PYZhy{\char`\-}

\def\PYZdq{\char`\"}

\makeatother

\providecommand*{\DUrole}[2]{%
  \ifcsname DUrole#1\endcsname%
    \csname DUrole#1\endcsname{#2}%
  \else
    \ifcsname docutilsrole#1\endcsname%
      \csname docutilsrole#1\endcsname{#2}%
    \else%
      #2%
    \fi%
  \fi%
}

\providecommand*{\DUroletitlereference}[1]{\textsl{#1}}

\ifthenelse{\isundefined{\hypersetup}}{
  \usepackage[colorlinks=true,linkcolor=blue,urlcolor=blue]{hyperref}
  \urlstyle{same} 
}{}

\begin{document}
\newcounter{footnotecounter}\title{Simulating X-ray Observations with Python}\author{John A. ZuHone$^{\setcounter{footnotecounter}{1}\fnsymbol{footnotecounter}\setcounter{footnotecounter}{2}\fnsymbol{footnotecounter}}$%
          \setcounter{footnotecounter}{1}\thanks{\fnsymbol{footnotecounter} %
          Corresponding author: \protect\href{mailto:jzuhone@milkyway.gsfc.nasa.gov}{jzuhone@milkyway.gsfc.nasa.gov}}\setcounter{footnotecounter}{2}\thanks{\fnsymbol{footnotecounter} Astrophysics Science Division, Laboratory for High Energy Astrophysics, Code 662, NASA/Goddard Space Flight Center, Greenbelt, MD 20771}, Veronica Biffi$^{\setcounter{footnotecounter}{3}\fnsymbol{footnotecounter}}$\setcounter{footnotecounter}{3}\thanks{\fnsymbol{footnotecounter} SISSA - Scuola Internazionale Superiore di Studi Avanzati, Via Bonomea 265, 34136 Trieste, Italy}, Eric J. Hallman$^{\setcounter{footnotecounter}{4}\fnsymbol{footnotecounter}}$\setcounter{footnotecounter}{4}\thanks{\fnsymbol{footnotecounter} Center for Astrophysics and Space Astronomy, Department of Astrophysical \& Planetary Science, University of Colorado, Boulder, CO 80309}, Scott W. Randall$^{\setcounter{footnotecounter}{5}\fnsymbol{footnotecounter}}$\setcounter{footnotecounter}{5}\thanks{\fnsymbol{footnotecounter} Harvard-Smithsonian Center for Astrophysics, 60 Garden Street, Cambridge, MA 02138}, Adam R. Foster$^{\setcounter{footnotecounter}{5}\fnsymbol{footnotecounter}}$, Christian Schmid$^{\setcounter{footnotecounter}{6}\fnsymbol{footnotecounter}}$\setcounter{footnotecounter}{6}\thanks{\fnsymbol{footnotecounter} Dr. Karl Remeis-Sternwarte \& ECAP, Sternwartstr. 7, 96049 Bamberg, Germany}\thanks{%

          \noindent%
          Copyright\,\copyright\,2014 John A. ZuHone et al. This is an open-access article distributed under the terms of the Creative Commons Attribution License, which permits unrestricted use, distribution, and reproduction in any medium, provided the original author and source are credited.%
        }}\maketitle
          \renewcommand{\leftmark}{PROC. OF THE 13th PYTHON IN SCIENCE CONF. (SCIPY 2014)}
          \renewcommand{\rightmark}{SIMULATING X-RAY OBSERVATIONS WITH PYTHON}

\InputIfFileExists{page_numbers.tex}{}{}
\newcommand*{\docutilsroleref}{\ref}
\newcommand*{\docutilsrolelabel}{\label}
\begin{abstract}X-ray astronomy is an important tool in the astrophysicist's toolkit to investigate
high-energy astrophysical phenomena. Theoretical numerical simulations of astrophysical
sources are fully three-dimensional representations of physical quantities such as
density, temperature, and pressure, whereas astronomical observations are
two-dimensional projections of the emission generated via mechanisms dependent on these
quantities. To bridge the gap between simulations and observations, algorithms for
generating synthetic observations of simulated data have been developed. We present an
implementation of such an algorithm in the \texttt{yt} analysis software package. We describe
the underlying model for generating the X-ray photons, the important role that \texttt{yt}
and other Python packages play in its implementation, and present a detailed workable
example of the creation of simulated X-ray observations.\end{abstract}

\subsection{Introduction%
  \label{introduction}%
}

In the early 21st century, astronomy is truly a multi-wavelength enterprise. Ground and space-based instruments across the electromagnetic spectrum, from radio waves to gamma rays, provide the most complete picture of the various physical processes governing the evolution of astrophysical sources. In particular, X-ray astronomy probes high-energy processes in astrophysics, including high-temperature thermal plasmas (e.g., the solar wind, the intracluster medium) and relativistic cosmic rays (e.g., from active galactic nuclei). X-ray astronomy has a long and successful pedigree, with a number of observatories. These include \href{http://heasarc.gsfc.nasa.gov/docs/einstein/heao2.html}{\emph{Einstein}}, \href{http://science.nasa.gov/missions/rosat/}{\emph{ROSAT}}, \href{http://chandra.harvard.edu}{\emph{Chandra}}, \href{http://xmm.esac.esa.int/}{\emph{XMM-Newton}}, \href{http://www.isas.jaxa.jp/e/enterp/missions/suzaku/}{\emph{Suzaku}}, and \href{http://www.nustar.caltech.edu/}{\emph{NuSTAR}}, as well as upcoming missions such as \href{http://astro-h.isas.jaxa.jp/en/}{\emph{Astro-H}} and \href{http://www.the-athena-x-ray-observatory.eu/}{\emph{Athena}}.

An important distinguishing feature of X-ray astronomy from that of studies at longer wavelengths is that it is inherently \DUroletitlereference{discrete}, e.g., the numbers of photons per second that reach the detectors are small enough that the continuum approximation, valid for longer-wavelength photons such as those in the visible light, infrared, microwave, and radio bands, is invalid. Instead of images, the fundamental data products of X-ray astronomy are tables of individual photon positions, energies, and arrival times.

Due to modeling uncertainties, projection effects, and contaminating backgrounds, combining the insights from observations and numerical simulations is not necessarily straightforward. In contrast to simulations, where all of the physical quantities in 3 dimensions are completely known to the precision of the simulation algorithm, astronomical observations are by definition 2-D projections of 3-D sources along a given sight line, and the observed spectrum of emission is a complicated function of the fundamental physical properties (e.g., density, temperature, composition) of the source.

Such difficulties in bridging these two worlds have given rise to efforts to close the gap in the direction of the creation of synthetic observations from simulated data (see, e.g., \cite{Gardini04}, \cite{Nagai06}, \cite{ZuHone09}, and \cite{Heinz11} for recent examples). This involves the determination of the spectrum of the emission from the properties of the source, the projection of this emission along the chosen line of sight, and, in the case of X-ray (and $\gamma$-ray) astronomy, the generation of synthetic photon samples. These photons are then convolved with the instrumental responses and (if necessary) background effects are added. One implementation of such a procedure, \href{http://www.mpa-garching.mpg.de/~kdolag/Phox/}{\texttt{PHOX}}, was described in \cite{Biffi12} and \cite{Biffi13}, and used for the analysis of simulated galaxy clusters from smoothed-particle hydrodynamics (SPH) cosmological simulations. \texttt{PHOX} was originally implemented in C using outputs from the \texttt{Gadget} SPH code. \texttt{PHOX} takes the inputs of density, temperature, velocity, and metallicity from a 3D \texttt{Gadget} simulation, using them as inputs to create synthetic spectra (using spectral models from the X-ray spectral fitting package \href{http://heasarc.gsfc.nasa.gov/xanadu/xspec}{\texttt{XSPEC}}). Finally, \texttt{PHOX} uses these synthetic spectra convolved with instrument response functions to simulate samples of observed photons.

In this work, we describe an extension of this algorithm to outputs from other simulation codes. We developed the module \texttt{photon\_simulator}, an implementation of \texttt{PHOX} within the Python-based \href{http://yt-project.org}{\texttt{yt}} simulation analysis package. We outline the design of the \texttt{PHOX} algorithm, the specific advantages to implementing it in Python and \texttt{yt}, and provide a workable example of the generation of a synthetic X-ray observation from a simulation dataset.

\subsection{Model%
  \label{model}%
}

The overall model that underlies the \texttt{PHOX} algorithm may be split up into roughly three steps: first, constructing an original large sample of simulated photons for a given source, second, choosing a subset of these photons corresponding to parameters appropriate to an actual observation and projecting them onto the sky plane, and finally, applying instrumental responses for a given detector. We briefly describe each of these in turn.\begin{figure*}[]\noindent\makebox[\textwidth][c]{\includegraphics[scale=0.25]{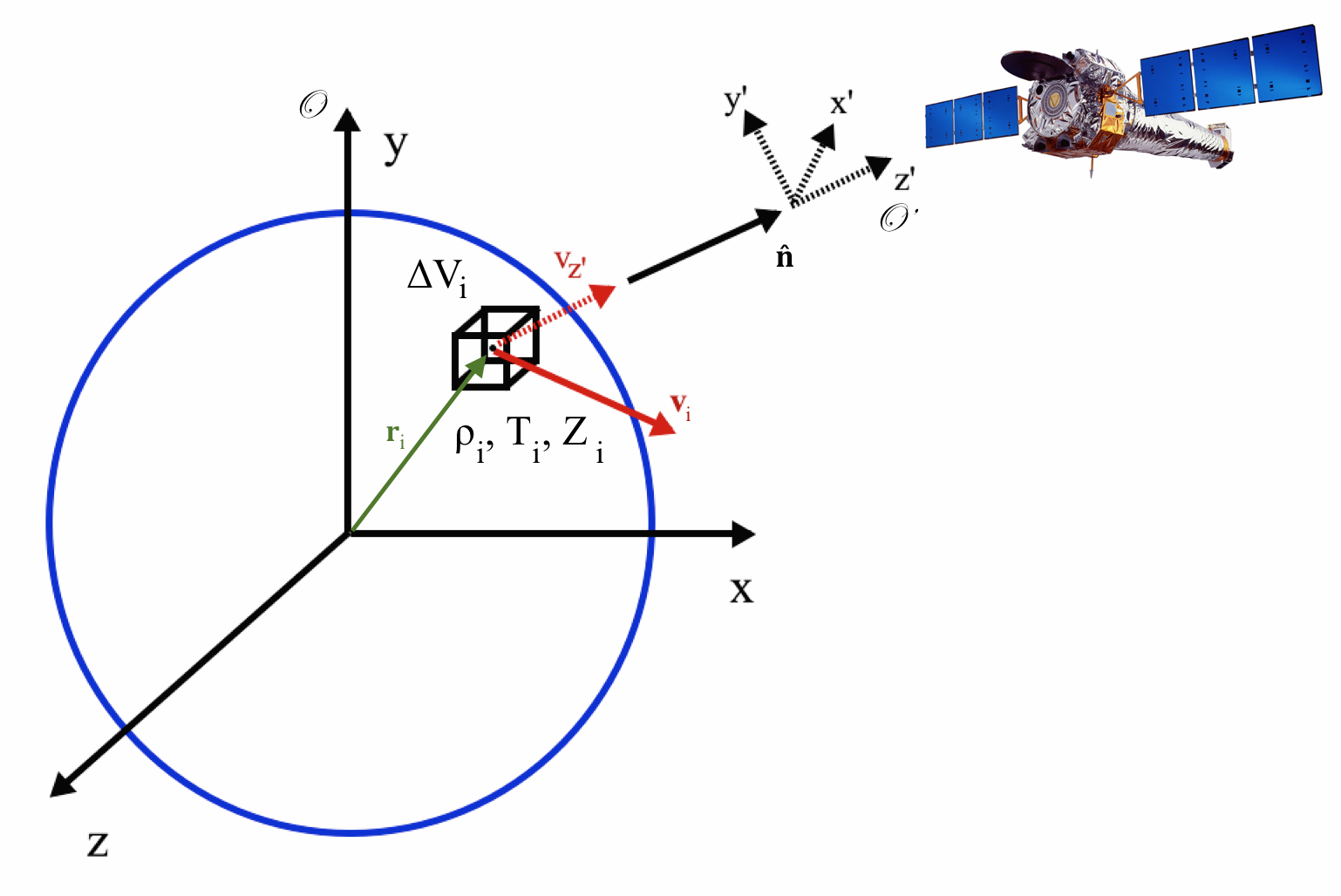}}
\caption{Schematic representation of a roughly spherical X-ray emitting object, such as a
galaxy cluster. The volume element $\Delta{V}_i$ at position ${\bf r}_i$
in the coordinate system ${\cal O}$ of the source has a velocity
${\bf v}_i$. Photons emitted along the direction given by $\hat{\bf n}$
will be received in the observer's frame in the coordinate system ${\cal O}'$,
and will be Doppler-shifted by the line-of-sight velocity component $v_{i,z'}$.
${\rm Chandra}$ telescope image credit: NASA/CXC. \DUrole{label}{schematic}}
\end{figure*}

\subsubsection{Step 1: Generating the Original Photon Sample%
  \label{step-1-generating-the-original-photon-sample}%
}

In the first step of the \texttt{PHOX} algorithm, we generate a large sample of photons in three dimensions, with energies in the rest frame of the source. These photons will serve as a \textquotedbl{}Monte-Carlo\textquotedbl{} sample from which we may draw subsets to construct realistic observations.

First, to determine the energies of the photons, a spectral model for the photon emissivity must be specified. In general, the normalization of the photon emissivity for a given volume element will be set by the number density of emitting particles, and the shape of the spectrum will be set by the energetics of the same particles.

As a specific and highly relevant example, one of the most common sources of X-ray emission is that from a low-density, high-temperature, thermal plasma, such as that found in the solar corona, supernova remnants, \textquotedbl{}early-type\textquotedbl{} galaxies, galaxy groups, and galaxy clusters. The specific photon count emissivity associated with a given density, temperature $T$, and metallicity $Z$ of such a plasma is given by\begin{equation}
\label{emissivity}
\epsilon_E^\gamma = n_en_H\Lambda_E(T,Z)~{\rm photons~s^{-1}~cm^{-3}~keV^{-1}}
\end{equation}where the superscript $\gamma$ refers to the fact that this is a photon count emissivity, $E$ is the photon energy in keV, $n_e$ and $n_H$ are the electron and proton number densities in ${\rm cm^{-3}}$, and $\Lambda_E(T,Z)$ is the spectral model in units of ${\rm photons~s^{-1}~cm^{3}~keV^{-1}}$. The dominant contributions to $\Lambda_E$ for an optically-thin, fully-ionized plasma are bremmstrahlung (\textquotedbl{}free-free\textquotedbl{}) emission and collisional line excitation. A number of models for the emissivity of such a plasma have been developed, including Raymond-Smith \cite{Raymond77}, MeKaL \cite{Mewe95}, and APEC \cite{Smith01}. These models (and others) are all built into the \texttt{XSPEC} package, which includes a Python interface, \href{http://heasarc.gsfc.nasa.gov/xanadu/xspec/python/html/}{\texttt{PyXspec}}, which is a package we will use to supply the input spectral models to generate the photon energies.

The original \texttt{PHOX} algorithm only allowed for emission from variants of the APEC model for a thermal plasma. However, astrophysical X-ray emission arises from a variety of physical processes and sources, and in some cases multiple sources may be emitting from within the same volume. For example, cosmic-ray electrons in galaxy clusters produce a power-law spectrum of X-ray emission at high energies via inverse-Compton scattering of the cosmic microwave background. Recently, the detection of previously unidentified line emission, potentially from decaying sterile neutrinos, was made in stacked spectra of galaxy clusters \cite{Bulbul14}. The flexibility of our approach allows us to implement one or several models for the X-ray emission arising from a variety of physical processes as the situation requires.

Given a spectral model, for a given volume element $i$ with volume $\Delta{V}_i$ (which may be grid cells or Lagrangian particles), a spectrum of photons may be generated. The \emph{total number} of photons that are generated in our initial sample per volume element $i$ is determined by other factors. We determine the number of photons for each volume element by artificially inflating the parameters that determine the number of photons received by an observer to values that are large compared to more realistic values. The inflated Monte-Carlo sample should be large enough that realistic sized subsets from it are statistically representative. In the description that follows, parameters with subscript \textquotedbl{}0\textquotedbl{} indicate those with \textquotedbl{}inflated\textquotedbl{} values, whereas we will drop the subscripts in the second step when choosing more realistic values.

To begin with, the bolometric flux of photons received by the observer from the volume element $i$ is\begin{equation}
\label{flux}
F^{\gamma}_i = \frac{n_{e,i}n_{H,i}\Lambda(T_i,Z_i)\Delta{V}_i}{4\pi{D_{A,0}^2}(1+z_0)^2}~{\rm photons~s^{-1}~cm^{-2}}
\end{equation}where $z_0$ is the cosmological redshift and $D_{A,0}$ is the angular diameter distance to the source (if the source is nearby, $z_0 \approx 0$ and $D_{A,0}$ is simply the distance to the source). The physical quantities of interest are constant across the volume element. The total number of photons associated with this flux for an instrument with a collecting area $A_{\rm det,0}$ and an observation with exposure time $t_{\rm exp,0}$ is given by\begin{equation}
\label{n_phot}
N_{\rm phot} = t_{\rm exp,0}A_{\rm det,0}\displaystyle\sum_i{F^{\gamma}_i}
\end{equation}By setting $t_{\rm exp,0}$ and $A_{\rm det,0}$ to values that are much larger than those associated with typical exposure times and actual detector areas, and setting $z_0$ to a value that corresponds to a nearby source (thus ensuring $D_{A,0}$ is similarly small), we ensure that we create suitably large Monte-Carlo sample to draw subsets of photons for more realistic observational parameters. Figure \DUrole{ref}{schematic} shows a schematic representation of this model for a roughly spherical source of X-ray photons, such as a galaxy cluster.

\subsubsection{Step 2: Projecting Photons to Create Specific Observations%
  \label{step-2-projecting-photons-to-create-specific-observations}%
}

The second step in the \texttt{PHOX} algorithm involves using this large 3-D sample of photons to create 2-D projections of simulated events, where a subsample of photons from the original Monte-Carlo sample is selected.

First, we choose a line-of-sight vector $\hat{\bf n}$ to define the primed coordinate system from which the photon sky positions $(x',y')$ in the observer's coordinate system ${\cal O}'$ are determined (c.f. Figure \DUrole{ref}{schematic}). The total emission from any extended object as a function of position on the sky is a projection of the total emission along the line of sight, minus the emission that has been either absorbed or scattered out of the sight-line along the way. In the current state of our implementation, we assume that the source is optically thin to the photons, so they pass essentially unimpeded from the source to the observer (with the caveat that some photons are absorbed by Galactic foreground gas). This is appropriate for most X-ray sources of interest.

Next, we must take into account processes that affect on the photon energies. The first, occurring at the source itself, is Doppler shifting and broadening of spectral lines, which arises from bulk motion of the gas and turbulence. Each volume element has a velocity ${\bf v}_i$ in ${\cal O}$, and the component $v_{i,z'}$ of this velocity along the line of sight results in a Doppler shift of each photon's energy of\begin{equation}
\label{doppler}
E_1 = E_0\sqrt{\frac{c+v_{z'}}{c-v_{z'}}}
\end{equation}where $E_1$ and $E_0$ are the Doppler-shifted and rest-frame energies of the photon, respectively, and $c$ is the speed of light in vacuum. Second, since many X-ray sources are at cosmological distances, each photon is cosmologically redshifted, reducing its energy further by a factor of $1/(1+z)$ before being received in the observer's frame.

Since we are now simulating an actual observation, we choose more realistic values for the exposure time $t_{\rm exp}$ and detector area $A_{\rm det}$ than we did in the first step to determine the number of photons to use from the original Monte-Carlo sample. Similarly, we may also specify alternative values for the angular diameter distance $D_A$ and the cosmological redshift $z$, if desired. The fraction $f$ of the photons that will be used in the actual observation is then given by\begin{equation}
\label{fraction}
f = \frac{t_{\rm exp}}{t_{\rm exp,0}}\frac{A_{\rm det}}{A_{\rm det,0}}\frac{D_{A,0}^2}{D_A^2}\frac{(1+z_0)^3}{(1+z)^3}
\end{equation}where $f \leq 1$.

Before being received by the observer, a number of the photons, particularly on the softer end of the spectrum, are absorbed by foregrounds of hydrogen gas in the Milky Way Galaxy. The last operation that is applied in our implementation of the \texttt{PHOX} algorithm is to use a tabulated model for the absorption cross-section as a function of energy (examples include \texttt{wabs} \cite{Morrison83}, \texttt{phabs} \cite{Balucinska-Church92}, \texttt{tbabs} \cite{Wilms00}, all included in \texttt{XSPEC}) as an acceptance-rejection criterion for which photons will be retained in the final sample, e.g., which of them are actually received by the observer.

The advantage of the \texttt{PHOX} algorithm is that the two steps of generating the photons in the source frame and projecting them along a given line of sight are separated, so that the first step, which is the most computationally expensive, need only be done once for a given source, whereas the typically cheaper second step may be repeated many times for many different lines of sight, different instruments, and different exposure times.

\subsubsection{Step 3: Modeling Instrumental Effects%
  \label{step-3-modeling-instrumental-effects}%
}

Unfortunately, the data products of X-ray observations do not simply consist of the original sky positions and energies of the received photons. Spatially, the positions of the received photons on the detector are affected by a number of instrumental factors. These include vignetting, the layout of the CCD chips, and a typically spatially dependent point-spread function. Similarly, the photon energies are binned up by the detectors into a set of discrete energy channels, and there is typically not a simple one-to-one mapping between which channel a given photon ends up in and its original energy, but is instead represented by a non-diagonal response matrix. Finally, the \textquotedbl{}effective\textquotedbl{} collecting area of the telescope is also energy-dependent, and also varies with position on the detector. When performing analysis of X-ray data, the mapping between the detector channel and the photon energy is generally encapsulated in a \href{http://cxc.harvard.edu/ciao/dictionary/rmf.html}{redistribution matrix file (RMF)}, and the effective area curve as a function of energy is encapsulated in an \href{http://cxc.harvard.edu/ciao/dictionary/arf.html}{ancillary response file} (ARF).

In our framework, we provide two ways of convolving the detected photons with instrumental responses, depending on the level of sophistication required. The first is a \textquotedbl{}bare-bones\textquotedbl{} approach, where the photon positions are convolved with a user-specified point-spread function, and the photon energies are convolved with a user-input energy response functions. This will result in photon distributions that are similar enough to the final data products of real observations to be sufficient for most purposes.

However, some users may require a full simulation of a given telescope or may wish to compare observations of the same simulated system by multiple instruments. Several software packages exist for this purpose. The venerable \href{http://space.mit.edu/ASC/MARX/}{\texttt{MARX}} software performs detailed ray-trace simulations of how \DUroletitlereference{Chandra} responds to a variety of astrophysical sources, and produces standard event data files in the same FITS formats as standard \DUroletitlereference{Chandra} data products. \href{http://hea-www.harvard.edu/simx/}{\texttt{SIMX}} and \href{http://www.sternwarte.uni-erlangen.de/research/sixte/}{\texttt{Sixte}} are similar packages that simulate most of the effects of the instrumental responses for a variety of current and planned X-ray missions. We provide convenient output formats for the synthetic photons in order that they may be easily imported into these packages.\begin{figure*}[]\noindent\makebox[\textwidth][c]{\includegraphics[]{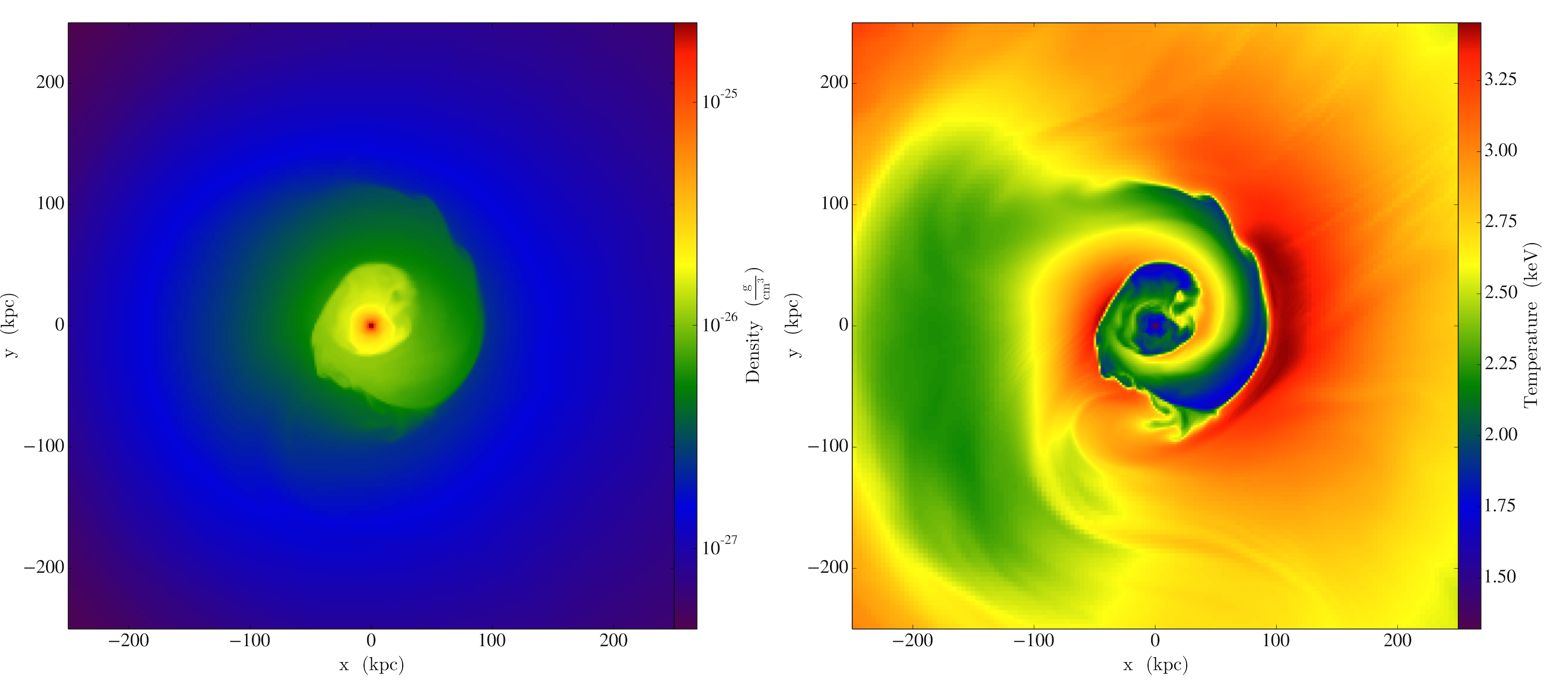}}
\caption{Slices of density (left) and temperature (right) of an \texttt{Athena} dataset of a
galaxy cluster core. \DUrole{label}{sloshing}}
\end{figure*}

\subsection{Implementation%
  \label{implementation}%
}

The model described here has been implemented as the analysis module \texttt{photon\_simulator} in \texttt{yt} \cite{Turk11}, a Python-based visualization and analysis toolkit for volumetric data. \texttt{yt} has a number of strengths that make it an ideal package for implementing our algorithm.

The first is that \texttt{yt} has support for analyzing data from a large number of astrophysical simulation codes (e.g., \href{http://flash.uchicago.edu}{\texttt{FLASH}}, \href{http://www.enzo-project.org}{\texttt{Enzo}}, \href{http://www.mpa-garching.mpg.de/gadget/}{\texttt{Gadget}}, \href{http://www.astro.princeton.edu/~jstone/athena.html}{\texttt{Athena}}), which simulate the formation and evolution of astrophysical systems using models for the relevant physics, including magnetohydrodynamics, gravity, dark matter, plasmas, etc. The simulation-specific code is contained within various \textquotedbl{}frontend\textquotedbl{} implementations, and the user-facing API to perform the analysis on the data is the same regardless of the type of simulation being analyzed. This enables the same function calls to easily generate photons from models produced by any of these simulation codes making it possible to use the \texttt{PHOX} algorithm beyond the original application to \texttt{Gadget} simulations only. In fact, most previous approaches to simulating X-ray observations were limited to datasets from particular simulation codes.

The second strength is related, in that by largely abstracting out the simulation-specific concepts of \textquotedbl{}cells\textquotedbl{}, \textquotedbl{}grids\textquotedbl{}, \textquotedbl{}particles\textquotedbl{}, \textquotedbl{}smoothing lengths\textquotedbl{}, etc., \texttt{yt} provides a window on to the data defined primarily in terms of physically motivated volumetric region objects. These include spheres, disks, rectangular regions, regions defined on particular cuts on fields, etc. Arbitrary combinations of these region types are also possible. These volumetric region objects serve as natural starting points for generating X-ray photons from not only physically relevant regions within a complex hydrodynamical simulation, but also from simple \textquotedbl{}toy\textquotedbl{} models which have been constructed from scratch, when complex, expensive simulations are not necessary.

The third major strength is that implementing our model in \texttt{yt} makes it possible to easily make use of the wide variety of useful libraries available within the scientific Python ecosystem. Our implementation uses \href{http://www.scipy.org}{\texttt{SciPy}} for integration, \href{http://www.astropy.org}{\texttt{AstroPy}} for handling celestial coordinate systems and FITS I/O, and \texttt{PyXspec} for generating X-ray spectral models. Tools for analyzing astrophysical X-ray data are also implemented in Python (e.g., \href{http://cxc.harvard.edu/ciao/}{\texttt{CIAO}}'s \href{http://cxc.harvard.edu/sherpa/}{\texttt{Sherpa}} package, \cite{Refsdal09}), enabling an easy comparison between models and observations.

\subsection{Example%
  \label{example}%
}

Here we present a workable example of creating simulated X-ray events using \texttt{yt}'s \texttt{photon\_simulator} analysis module. We implemented the module in \texttt{yt} v. 3.0 as \texttt{yt.analysis\_modules.photon\_simulator}. \texttt{yt} v. 3.0 can be downloaded from \url{http://yt-project.org}. The example code here is available \href{http://nbviewer.ipython.org/url/www.jzuhone.com/files/photon_simulator_example.ipynb}{as an IPython notebook}. This is not meant to be an exhaustive explanation of all of the \texttt{photon\_simulator}'s features and options-{}-for these the reader is encouraged to visit the \href{http://yt-project.org/doc/}{yt documentation}.

As our input dataset, we will use an \texttt{Athena} simulation of a galaxy cluster core, which can be downloaded from the \texttt{yt} website at \url{http://yt-project.org/data/MHDSloshing.tar.gz}.
You will also need to download a version of \texttt{APEC} from \url{http://www.atomdb.org}. Finally, the absorption cross section table used here and the \emph{Chandra} response files may be downloaded from \url{http://yt-project.org/data/xray_data.tar.gz}.

First, we must import the necessary modules:\vspace{1mm}
\begin{Verbatim}[commandchars=\\\{\},fontsize=\footnotesize]
\PY{k+kn}{import} \PY{n+nn}{yt}
\PY{k+kn}{from} \PY{n+nn}{yt.analysis\PYZus{}modules.photon\PYZus{}simulator.api} \PYZbs{}
    \PY{k+kn}{import} \PY{n}{TableApecModel}\PY{p}{,} \PY{n}{ThermalPhotonModel}\PY{p}{,} \PYZbs{}
    \PY{n}{PhotonList}\PY{p}{,} \PY{n}{TableAbsorbModel}
\PY{k+kn}{from} \PY{n+nn}{yt.utilities.cosmology} \PY{k+kn}{import} \PY{n}{Cosmology}
\end{Verbatim}
\vspace{1mm}
Next, we load the dataset \texttt{ds}, which comes from a set of simulations presented in \cite{ZuHone14}. \texttt{Athena} datasets require a \texttt{parameters} dictionary to be supplied to provide unit conversions to Gaussian units; for most datasets generated by other simulation codes that can be read by \texttt{yt}, this is not necessary.\vspace{1mm}
\begin{Verbatim}[commandchars=\\\{\},fontsize=\footnotesize]
\PY{n}{parameters}\PY{o}{=}\PY{p}{\PYZob{}}\PY{l+s}{\PYZdq{}}\PY{l+s}{time\PYZus{}unit}\PY{l+s}{\PYZdq{}}\PY{p}{:}\PY{p}{(}\PY{l+m+mf}{1.0}\PY{p}{,}\PY{l+s}{\PYZdq{}}\PY{l+s}{Myr}\PY{l+s}{\PYZdq{}}\PY{p}{)}\PY{p}{,}
            \PY{l+s}{\PYZdq{}}\PY{l+s}{length\PYZus{}unit}\PY{l+s}{\PYZdq{}}\PY{p}{:}\PY{p}{(}\PY{l+m+mf}{1.0}\PY{p}{,}\PY{l+s}{\PYZdq{}}\PY{l+s}{Mpc}\PY{l+s}{\PYZdq{}}\PY{p}{)}\PY{p}{,}
            \PY{l+s}{\PYZdq{}}\PY{l+s}{mass\PYZus{}unit}\PY{l+s}{\PYZdq{}}\PY{p}{:}\PY{p}{(}\PY{l+m+mf}{1.0e14}\PY{p}{,}\PY{l+s}{\PYZdq{}}\PY{l+s}{Msun}\PY{l+s}{\PYZdq{}}\PY{p}{)}\PY{p}{\PYZcb{}}

\PY{n}{ds} \PY{o}{=} \PY{n}{yt}\PY{o}{.}\PY{n}{load}\PY{p}{(}\PY{l+s}{\PYZdq{}}\PY{l+s}{MHDSloshing/virgo\PYZus{}low\PYZus{}res.0054.vtk}\PY{l+s}{\PYZdq{}}\PY{p}{,}
             \PY{n}{parameters}\PY{o}{=}\PY{n}{parameters}\PY{p}{)}
\end{Verbatim}
\vspace{1mm}
Slices through the density and temperature of the simulation dataset are shown in Figure \DUrole{ref}{sloshing}. The luminosity and temperature of our model galaxy cluster roughly match that of Virgo. The photons will be created from a spherical region centered on the domain center, with a radius of 250 kpc:\vspace{1mm}
\begin{Verbatim}[commandchars=\\\{\},fontsize=\footnotesize]
\PY{n}{sp} \PY{o}{=} \PY{n}{ds}\PY{o}{.}\PY{n}{sphere}\PY{p}{(}\PY{l+s}{\PYZdq{}}\PY{l+s}{c}\PY{l+s}{\PYZdq{}}\PY{p}{,} \PY{p}{(}\PY{l+m+mf}{250.}\PY{p}{,} \PY{l+s}{\PYZdq{}}\PY{l+s}{kpc}\PY{l+s}{\PYZdq{}}\PY{p}{)}\PY{p}{)}
\end{Verbatim}
\vspace{1mm}
This will serve as our \texttt{data\_source} that we will use later. Now, we are ready to use the \texttt{photon\_simulator} analysis module to create synthetic X-ray photons from this dataset.

\subsubsection{Step 1: Generating the Original Photon Sample%
  \label{id17}%
}

First, we need to create the \texttt{SpectralModel} instance that will determine how
the data in the grid cells will generate photons. A number of options are available, but we will use the \texttt{TableApecModel}, which allows one to use the \texttt{APEC} data tables:\vspace{1mm}
\begin{Verbatim}[commandchars=\\\{\},fontsize=\footnotesize]
\PY{n}{atomdb\PYZus{}path} \PY{o}{=} \PY{l+s}{\PYZdq{}}\PY{l+s}{/Users/jzuhone/Data/atomdb}\PY{l+s}{\PYZdq{}}

\PY{n}{apec\PYZus{}model} \PY{o}{=} \PY{n}{TableApecModel}\PY{p}{(}\PY{n}{atomdb\PYZus{}path}\PY{p}{,}
                            \PY{l+m+mf}{0.01}\PY{p}{,} \PY{l+m+mf}{10.0}\PY{p}{,} \PY{l+m+mi}{2000}\PY{p}{,}
                            \PY{n}{apec\PYZus{}vers}\PY{o}{=}\PY{l+s}{\PYZdq{}}\PY{l+s}{2.0.2}\PY{l+s}{\PYZdq{}}\PY{p}{,}
                            \PY{n}{thermal\PYZus{}broad}\PY{o}{=}\PY{n+nb+bp}{False}\PY{p}{)}
\end{Verbatim}
\vspace{1mm}
where the first argument specifies the path to the \texttt{APEC} files, the next three specify the bounds in keV of the energy spectrum and the number of bins in the table, and the remaining arguments specify the \texttt{APEC} version to use and whether or not to apply thermal broadening to the spectral lines.

Now that we have our \texttt{SpectralModel}, we need to connect this model to a \texttt{PhotonModel} that will connect the field data in the \texttt{data\_source} to the spectral model to and generate the photons which will serve as the sample distribution for observations. For thermal spectra, we have a special \texttt{PhotonModel} called \texttt{ThermalPhotonModel}:\vspace{1mm}
\begin{Verbatim}[commandchars=\\\{\},fontsize=\footnotesize]
\PY{n}{thermal\PYZus{}model} \PY{o}{=} \PY{n}{ThermalPhotonModel}\PY{p}{(}\PY{n}{apec\PYZus{}model}\PY{p}{,}
                                   \PY{n}{X\PYZus{}H}\PY{o}{=}\PY{l+m+mf}{0.75}\PY{p}{,}
                                   \PY{n}{Zmet}\PY{o}{=}\PY{l+m+mf}{0.3}\PY{p}{)}
\end{Verbatim}
\vspace{1mm}
Where we pass in the \texttt{SpectralModel}, and can optionally set values for
the hydrogen mass fraction \texttt{X\_H} and metallicity \texttt{Z\_met}, the latter of which may be a single floating-point value or the name of the \texttt{yt} field representing the spatially-dependent metallicity.

Next, we need to specify \textquotedbl{}fiducial\textquotedbl{} values for the telescope collecting area in ${\rm cm}^2$, exposure time in seconds, and cosmological redshift, choosing generous values so that there will be a large number of photons in the Monte-Carlo sample. We also construct a \texttt{Cosmology} object, which will be used to determine the source distance from its redshift.\vspace{1mm}
\begin{Verbatim}[commandchars=\\\{\},fontsize=\footnotesize]
\PY{n}{A} \PY{o}{=} \PY{l+m+mf}{6000.} \PY{c}{\PYZsh{} must be in cm**2!}
\PY{n}{exp\PYZus{}time} \PY{o}{=} \PY{l+m+mf}{4.0e5} \PY{c}{\PYZsh{} must be in seconds!}
\PY{n}{redshift} \PY{o}{=} \PY{l+m+mf}{0.05}
\PY{n}{cosmo} \PY{o}{=} \PY{n}{Cosmology}\PY{p}{(}\PY{p}{)}
\end{Verbatim}
\vspace{1mm}
By default the \texttt{Cosmology} object uses the WMAP7 cosmological parameters from \cite{Komatsu11}, but others may be supplied, such as the \cite{Planck13} parameters:\vspace{1mm}
\begin{Verbatim}[commandchars=\\\{\},fontsize=\footnotesize]
\PY{n}{cosmo} \PY{o}{=} \PY{n}{Cosmology}\PY{p}{(}\PY{n}{hubble\PYZus{}constant} \PY{o}{=} \PY{l+m+mf}{0.67}\PY{p}{,}
                  \PY{n}{omega\PYZus{}matter} \PY{o}{=} \PY{l+m+mf}{0.32}\PY{p}{,}
                  \PY{n}{omega\PYZus{}lambda} \PY{o}{=} \PY{l+m+mf}{0.68}\PY{p}{)}
\end{Verbatim}
\vspace{1mm}
Now, we finally combine everything together and create a \texttt{PhotonList}
instance, which contains the photon samples:\vspace{1mm}
\begin{Verbatim}[commandchars=\\\{\},fontsize=\footnotesize]
\PY{n}{photons} \PY{o}{=} \PY{n}{PhotonList}\PY{o}{.}\PY{n}{from\PYZus{}scratch}\PY{p}{(}\PY{n}{sp}\PY{p}{,} \PY{n}{redshift}\PY{p}{,} \PY{n}{A}\PY{p}{,}
                                  \PY{n}{exp\PYZus{}time}\PY{p}{,}
                                  \PY{n}{thermal\PYZus{}model}\PY{p}{,}
                                  \PY{n}{center}\PY{o}{=}\PY{l+s}{\PYZdq{}}\PY{l+s}{c}\PY{l+s}{\PYZdq{}}\PY{p}{,}
                                  \PY{n}{cosmology}\PY{o}{=}\PY{n}{cosmo}\PY{p}{)}
\end{Verbatim}
\vspace{1mm}
where we have used all of the parameters defined above, and \texttt{center} defines the reference coordinate which will become the origin of the photon coordinates, which in this case is \texttt{\textquotedbl{}c\textquotedbl{}}, the center of the simulation domain. This object contains the positions and velocities of the originating volume elements of the photons, as well as their rest-frame energies.

Generating this Monte-Carlo sample is the most computationally intensive part of the PHOX algorithm. Once a sample has been generated it can be saved to disk and loaded as needed rather than needing to be regenerated for different observational scenarios (instruments, redshifts, etc). The photons object can be saved to disk in the \href{http://www.hdfgroup.org}{HDF5} format with the following method:\vspace{1mm}
\begin{Verbatim}[commandchars=\\\{\},fontsize=\footnotesize]
\PY{n}{photons}\PY{o}{.}\PY{n}{write\PYZus{}h5\PYZus{}file}\PY{p}{(}\PY{l+s}{\PYZdq{}}\PY{l+s}{my\PYZus{}photons.h5}\PY{l+s}{\PYZdq{}}\PY{p}{)}
\end{Verbatim}
\vspace{1mm}
To load these photons at a later time, we use the \texttt{from\_file} method:\vspace{1mm}
\begin{Verbatim}[commandchars=\\\{\},fontsize=\footnotesize]
\PY{n}{photons} \PY{o}{=} \PY{n}{PhotonList}\PY{o}{.}\PY{n}{from\PYZus{}file}\PY{p}{(}\PY{l+s}{\PYZdq{}}\PY{l+s}{my\PYZus{}photons.h5}\PY{l+s}{\PYZdq{}}\PY{p}{)}
\end{Verbatim}
\vspace{1mm}

\subsubsection{Step 2: Projecting Photons to Create Specific Observations%
  \label{id20}%
}
At this point the photons can be projected along a line of sight to create a specific synthetic observation. First, it is necessary to set up a spectral model for the Galactic absorption cross-section, similar to the spectral model for the emitted photons set up previously. Here again, there are multiple options, but for the current example we use \texttt{TableAbsorbModel}, which allows one to use an absorption cross section vs. energy table written in HDF5 format (available in the \href{http://yt-project.org/data/xray_data.tar.gz}{xray\_data.tar.gz} file mentioned previously). This method also takes the column density \texttt{N\_H} in units of $10^{22}~{\rm cm}^{-2}$ as an additional argument.\vspace{1mm}
\begin{Verbatim}[commandchars=\\\{\},fontsize=\footnotesize]
\PY{n}{N\PYZus{}H} \PY{o}{=} \PY{l+m+mf}{0.1}
\PY{n}{a\PYZus{}mod} \PY{o}{=} \PY{n}{TableAbsorbModel}\PY{p}{(}\PY{l+s}{\PYZdq{}}\PY{l+s}{tbabs\PYZus{}table.h5}\PY{l+s}{\PYZdq{}}\PY{p}{,} \PY{n}{N\PYZus{}H}\PY{p}{)}
\end{Verbatim}
\vspace{1mm}
We next set a line-of-sight vector \texttt{L}:\vspace{1mm}
\begin{Verbatim}[commandchars=\\\{\},fontsize=\footnotesize]
\PY{n}{L} \PY{o}{=} \PY{p}{[}\PY{l+m+mf}{0.0}\PY{p}{,} \PY{l+m+mf}{0.0}\PY{p}{,} \PY{l+m+mf}{1.0}\PY{p}{]}
\end{Verbatim}
\vspace{1mm}
which corresponds to the direction within the simulation domain along which the photons will be projected. The exposure time, telescope area, and source redshift may also be optionally set to more appropriate values for a particular observation:\vspace{1mm}
\begin{Verbatim}[commandchars=\\\{\},fontsize=\footnotesize]
\PY{n}{texp} \PY{o}{=} \PY{l+m+mf}{1.0e5}
\PY{n}{z} \PY{o}{=} \PY{l+m+mf}{0.07}
\end{Verbatim}
\vspace{1mm}
If any of them are not set, those parameters will be set to the original values used when creating the \texttt{photons} object.

Finally, an \texttt{events} object is created using the line-of-sight vector, modified observation parameters, and the absorption model:\vspace{1mm}
\begin{Verbatim}[commandchars=\\\{\},fontsize=\footnotesize]
\PY{n}{events} \PY{o}{=} \PY{n}{photons}\PY{o}{.}\PY{n}{project\PYZus{}photons}\PY{p}{(}\PY{n}{L}\PY{p}{,}
                                 \PY{n}{exp\PYZus{}time\PYZus{}new}\PY{o}{=}\PY{n}{texp}\PY{p}{,}
                                 \PY{n}{redshift\PYZus{}new}\PY{o}{=}\PY{n}{z}\PY{p}{,}
                                 \PY{n}{absorb\PYZus{}model}\PY{o}{=}\PY{n}{a\PYZus{}mod}\PY{p}{)}
\end{Verbatim}
\vspace{1mm}
\texttt{project\_photons} draws events uniformly from the \texttt{photons} sample, orients their positions in the coordinate frame defined by \texttt{L}, and applies the Doppler and cosmological energy shifts, and removes a number of events corresponding to the supplied Galactic absorption model.\begin{figure}[]\noindent\makebox[\columnwidth][c]{\includegraphics[scale=0.33]{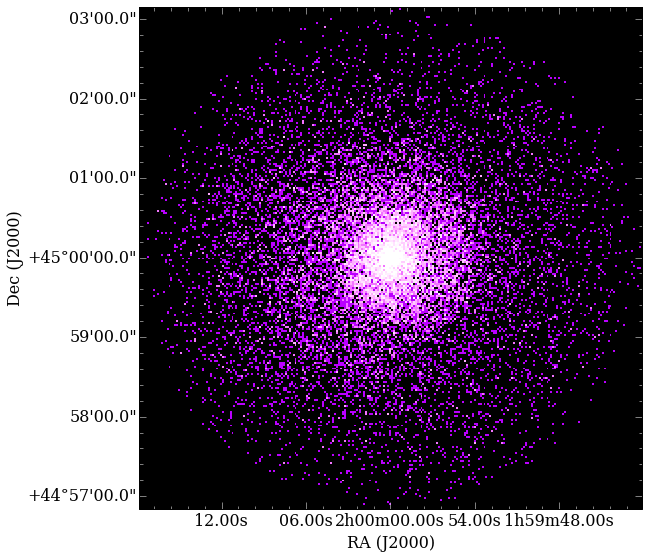}}
\caption{100 ks exposure of our simulated galaxy cluster, from a FITS image plotted with
\texttt{APLpy}. \DUrole{label}{image}}
\end{figure}

\subsubsection{Step 3: Modeling Instrumental Effects%
  \label{id21}%
}

If desired, instrumental response functions may be supplied to convolve the photons with a particular instrumental model. The files containing these functions are defined and put in a single list \texttt{resp}:\vspace{1mm}
\begin{Verbatim}[commandchars=\\\{\},fontsize=\footnotesize]
\PY{n}{ARF} \PY{o}{=} \PY{l+s}{\PYZdq{}}\PY{l+s}{chandra\PYZus{}ACIS\PYZhy{}S3\PYZus{}onaxis\PYZus{}arf.fits}\PY{l+s}{\PYZdq{}}
\PY{n}{RMF} \PY{o}{=} \PY{l+s}{\PYZdq{}}\PY{l+s}{chandra\PYZus{}ACIS\PYZhy{}S3\PYZus{}onaxis\PYZus{}rmf.fits}\PY{l+s}{\PYZdq{}}
\PY{n}{resp} \PY{o}{=} \PY{p}{[}\PY{n}{ARF}\PY{p}{,}\PY{n}{RMF}\PY{p}{]}
\end{Verbatim}
\vspace{1mm}
In this case, we would replace our previous call to \texttt{project\_photons} with one that supplies \texttt{resp} as the \texttt{responses} argument:\vspace{1mm}
\begin{Verbatim}[commandchars=\\\{\},fontsize=\footnotesize]
\PY{n}{events} \PY{o}{=} \PY{n}{photons}\PY{o}{.}\PY{n}{project\PYZus{}photons}\PY{p}{(}\PY{n}{L}\PY{p}{,}
                                 \PY{n}{exp\PYZus{}time\PYZus{}new}\PY{o}{=}\PY{n}{texp}\PY{p}{,}
                                 \PY{n}{redshift\PYZus{}new}\PY{o}{=}\PY{n}{z}\PY{p}{,}
                                 \PY{n}{absorb\PYZus{}model}\PY{o}{=}\PY{n}{a\PYZus{}mod}\PY{p}{,}
                                 \PY{n}{responses}\PY{o}{=}\PY{n}{resp}\PY{p}{)}
\end{Verbatim}
\vspace{1mm}
Supplying instrumental responses is optional. If they are provided, \texttt{project\_photons} performs 2 additional calculations. If an ARF is provided, the maximum value of the effective area curve will serve as the \texttt{area\_new} parameter, and after the absorption step a number of events are further removed using the effective area curve as the acceptance/rejection criterion. If an RMF is provided, it will be convolved with the event energies to produce a new array with the resulting spectral channels.\begin{figure}[]\noindent\makebox[\columnwidth][c]{\includegraphics[scale=0.33]{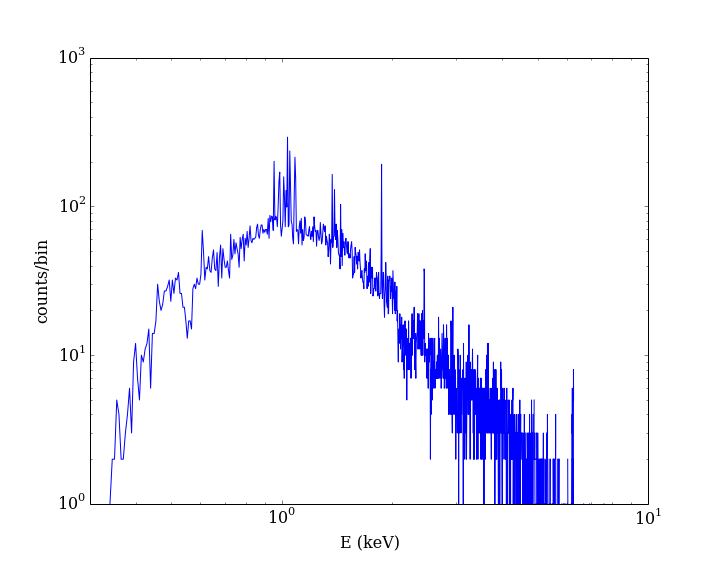}}
\caption{Spectral energy distribution of our simulated observation. \DUrole{label}{spectrum}}
\end{figure}\begin{figure*}[]\noindent\makebox[\textwidth][c]{\includegraphics[scale=0.50]{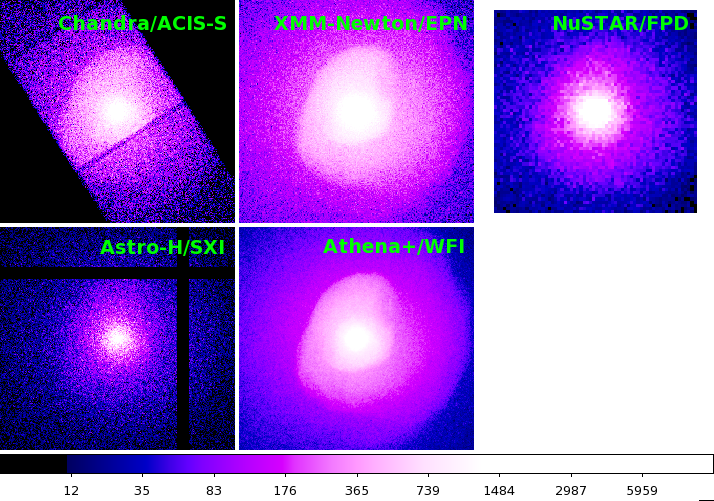}}
\caption{100 ks exposures of our simulated galaxy cluster, observed with several
different existing and planned X-ray detectors. The \DUroletitlereference{Chandra} image
was made with \texttt{MARX}, while the others were made with \texttt{SIMX}. All images have the
same angular scale. \DUrole{label}{comparison}}
\end{figure*}

However, if a more accurate simulation of a particular X-ray instrument is needed, or if one wishes to simulate multiple instruments, there are a couple of options for outputting our simulated events to be used by other software that performs such simulations. Since these external packages apply instrument response functions to the events list, the original \texttt{events} object generated from the \texttt{project\_photons} method must not be convolved with instrument responses (e.g., the ARF and RMF) in that step. For input to \texttt{MARX}, we provide an implementation of a \texttt{MARX} \textquotedbl{}user source\textquotedbl{} at \url{http://bitbucket.org/jzuhone/yt_marx_source}, which takes as input an HDF5 file. The events list can be written in the HDF5 file format with the following method:\vspace{1mm}
\begin{Verbatim}[commandchars=\\\{\},fontsize=\footnotesize]
\PY{n}{events}\PY{o}{.}\PY{n}{write\PYZus{}h5\PYZus{}file}\PY{p}{(}\PY{l+s}{\PYZdq{}}\PY{l+s}{my\PYZus{}events.h5}\PY{l+s}{\PYZdq{}}\PY{p}{)}
\end{Verbatim}
\vspace{1mm}
Input to \texttt{SIMX} and \texttt{Sixte} is handled via \href{http://hea-www.harvard.edu/heasarc/formats/simput-1.0.0.pdf}{\texttt{SIMPUT}}, a file format designed specifically for the output of simulated X-ray data. The events list can be written in SIMPUT file format with the following method:\vspace{1mm}
\begin{Verbatim}[commandchars=\\\{\},fontsize=\footnotesize]
\PY{n}{events}\PY{o}{.}\PY{n}{write\PYZus{}simput\PYZus{}file}\PY{p}{(}\PY{l+s}{\PYZdq{}}\PY{l+s}{my\PYZus{}events}\PY{l+s}{\PYZdq{}}\PY{p}{,}
                         \PY{n}{clobber}\PY{o}{=}\PY{n+nb+bp}{True}\PY{p}{,}
                         \PY{n}{emin}\PY{o}{=}\PY{l+m+mf}{0.1}\PY{p}{,} \PY{n}{emax}\PY{o}{=}\PY{l+m+mf}{10.0}\PY{p}{)}
\end{Verbatim}
\vspace{1mm}
where \texttt{emin} and \texttt{emax} are the energy range in keV of the outputted events. Figure \DUrole{ref}{comparison} shows several examples of the generated photons passed through various instrument simulations. \texttt{SIMX} and \texttt{MARX} produce FITS event files that are the same format as the data products of the individual telescope pipelines, so they can be analyzed by the same tools as real observations (e.g., \texttt{XSPEC}, \texttt{CIAO}).

\subsubsection{Examining the Data%
  \label{examining-the-data}%
}

The \texttt{events} may be binned into an image and written to a FITS file:\vspace{1mm}
\begin{Verbatim}[commandchars=\\\{\},fontsize=\footnotesize]
\PY{n}{events}\PY{o}{.}\PY{n}{write\PYZus{}fits\PYZus{}image}\PY{p}{(}\PY{l+s}{\PYZdq{}}\PY{l+s}{my\PYZus{}image.fits}\PY{l+s}{\PYZdq{}}\PY{p}{,}
                        \PY{n}{clobber}\PY{o}{=}\PY{n+nb+bp}{True}\PY{p}{,}
                        \PY{n}{emin}\PY{o}{=}\PY{l+m+mf}{0.5}\PY{p}{,} \PY{n}{emax}\PY{o}{=}\PY{l+m+mf}{7.0}\PY{p}{)}
\end{Verbatim}
\vspace{1mm}
where \texttt{emin} and \texttt{emax} specify the energy range for the image. Figure \DUrole{ref}{image} shows the resulting FITS image plotted using \href{http://aplpy.github.io/}{\texttt{APLpy}}.

We can also create a spectral energy distribution (SED) by binning the spectrum into energy bins. The resulting SED can be saved as a FITS binary table using the \texttt{write\_spectrum} method. In this example we bin up the spectrum according to the original photon energy, before it was convolved with the instrumental responses:\vspace{1mm}
\begin{Verbatim}[commandchars=\\\{\},fontsize=\footnotesize]
\PY{n}{events}\PY{o}{.}\PY{n}{write\PYZus{}spectrum}\PY{p}{(}\PY{l+s}{\PYZdq{}}\PY{l+s}{my\PYZus{}spec.fits}\PY{l+s}{\PYZdq{}}\PY{p}{,}
                      \PY{n}{energy\PYZus{}bins}\PY{o}{=}\PY{n+nb+bp}{True}\PY{p}{,}
                      \PY{n}{emin}\PY{o}{=}\PY{l+m+mf}{0.1}\PY{p}{,} \PY{n}{emax}\PY{o}{=}\PY{l+m+mf}{10.0}\PY{p}{,}
                      \PY{n}{nchan}\PY{o}{=}\PY{l+m+mi}{2000}\PY{p}{,} \PY{n}{clobber}\PY{o}{=}\PY{n+nb+bp}{True}\PY{p}{)}
\end{Verbatim}
\vspace{1mm}
here \texttt{energy\_bins} specifies whether we want to bin the events in unconvolved photon energy or convolved photon channel. Figure \DUrole{ref}{spectrum} shows the resulting spectrum.

\subsection{Summary%
  \label{summary}%
}

We have developed an analysis module within the Python-based volumetric data analysis toolkit \texttt{yt} to construct synthetic X-ray observations of astrophysical sources from simulation datasets, based on the \texttt{PHOX} algorithm. This algorithm generates a large sample of X-ray photons in the rest frame of the source from the physical quantities of the simulation dataset, and uses these as a sample from which a smaller number of photons are drawn and projected onto the sky plane, to simulate observations with a real detector. The utility of this algorithm lies in the fact that the most expensive step, namely that of generating the photons from the source, need only be done once, and these may be used as a Monte Carlo sample from which to generate as many simulated observations along as many projections and with as many instrument models as desired.

We implement PHOX in Python, using \texttt{yt} as an interface to the underlying simulation dataset. Our implementation takes advantage of the full range of capabilities of \texttt{yt}, especially its focus on physically motivated representations of simulation data and its support for a wide variety of simulation codes as well as generic \texttt{NumPy} array data generated on-the-fly. We also benefit from the object-oriented capabilities of Python as well as the ability to interface with existing astronomical and scientific Python packages.

Our module provides a crucial link between observations of astronomical sources and the simulations designed to represent the objects that are detected via their electromagnetic radiation, enabling some of the most direct testing of these simulations. Also, it is useful as a proposer's tool, allowing observers to generate simulated observations of astrophysical systems, to precisely quantify and motivate the needs of a proposal for observing time on a particular instrument. Our software also serves as a model for how similar modules in other wavebands may be designed, particularly in its use of several important Python packages for astronomy.

\end{document}